\documentclass[10pt,twocolumn,english,british,tightenlines,eqsecnum,
floats,aps,amsmath,amssymb,nofootinbib,superscriptaddress,prd,showpacs,showkeys]
              {revtex4}
\usepackage[latin9]{inputenc}
\usepackage{color}
\usepackage{babel}
\usepackage{amsmath}
\usepackage{amssymb}
\usepackage{graphicx}
\usepackage{esint}
\usepackage[unicode=true,pdfusetitle,
 bookmarks=true,bookmarksnumbered=false,bookmarksopen=false,
 breaklinks=false,pdfborder={0 0 1},backref=false,colorlinks=false]
 {hyperref}

\makeatletter
\@ifundefined{textcolor}{}
{%
 \definecolor{BLACK}{gray}{0}
 \definecolor{WHITE}{gray}{1}
 \definecolor{RED}{rgb}{1,0,0}
 \definecolor{GREEN}{rgb}{0,1,0}
 \definecolor{BLUE}{rgb}{0,0,1}
 \definecolor{CYAN}{cmyk}{1,0,0,0}
 \definecolor{MAGENTA}{cmyk}{0,1,0,0}
 \definecolor{YELLOW}{cmyk}{0,0,1,0}
}

\@ifundefined{date}{}{\date{October 31, 2016}}
\usepackage{babel}

\usepackage{euscript}\usepackage{epsfig}

\usepackage{amsfonts}

\usepackage{fancyhdr}

\makeatother

\begin{document}

\title{Gravity-Driven Acceleration and Kinetic Inflation in Noncommutative Brans-Dicke Setting}

\author{S. M. M. Rasouli}

\email{mrasouli@ubi.pt}

\affiliation{Departamento de F\'{i}sica, Universidade da Beira Interior, Rua Marqu\^{e}s d'Avila
e Bolama, 6200 Covilh\~{a}, Portugal}

\affiliation{Centro de Matem\'{a}tica e Aplica\c{c}\~{o}es (CMA - UBI),
Universidade da Beira Interior, Rua Marqu\^{e}s d'Avila
e Bolama, 6200 Covilh\~{a}, Portugal}

\author{Paulo Vargas Moniz}

\email{pmoniz@ubi.pt}

\affiliation{Departamento de F\'{i}sica, Universidade da Beira Interior, Rua Marqu\^{e}s d'Avila
e Bolama, 6200 Covilh\~{a}, Portugal}

\affiliation{Centro de Matem\'{a}tica e Aplica\c{c}\~{o}es (CMA - UBI),
Universidade da Beira Interior, Rua Marqu\^{e}s d'Avila e Bolama, 6200 Covilh\~{a}, Portugal}

\begin{abstract}
By assuming the spatially flat~FLRW line-element and
employing the Hamiltonian formalism, a noncommutative (NC) setting of the Brans-Dicke (BD)
theory is introduced.
We investigate gravity-driven
acceleration and kinetic inflation in this NC BD cosmology.
Despite to the commutative case, in which both the scale factor and BD scalar
field are obtained in power-law forms (in terms of the cosmic time), in our herein  NC model, we
see that the power-law scalar factor is multiplied by a dynamical
exponential warp factor. This warp factor depends on not only the noncommutative
parameter but also the momentum conjugate associated to the BD scalar field.
For very small values of this parameter, we obtain
an appropriate inflationary solution, which can
overcome problems within standard BD cosmology in a more efficient manner.
Moreover, we see that a graceful exit from an early acceleration epoch towards a
decelerating radiation epoch is provided.
For late times, due to the presence of the NC parameter, we obtain a zero
acceleration epoch, which can be interpreted as the coarse-grained explanation.
 \end{abstract}

\medskip


\maketitle

\section{Introduction}
In the BD setting employed in~\cite{Lev95,Lev95-2},
a variable BD coupling parameter rather than a constant one has been supposed.
Then, without
introducing any scalar potential or cosmological constant, an accelerated
expanding universe, from
the kinetic
energy density of a dynamical Planck mass, has been obtained.
In~\cite{Lev95-2}, it has been argued that, to meet
sufficient inflation, the scale factor in the Einstein frame must accelerate.
However, there is no source to get an accelerating scale
factor in that frame.
In other words, in the commutative case of the BD
theory~\cite{BD61} (in the Jordan frame), even by assuming variable $\omega$, there
is a fundamental problem with kinetic
inflation:
regardless of the form of $\omega(\phi)$, all the
D branch\footnote{We will introduce the D and X branches later.}
solutions are encountered with the graceful exit problem~\cite{Lev95-2}.

In our herein NC BD setting\footnote{Most of the discussions of this work has been reported
in our previous paper~\cite{RM14}.}, which, regardless of varying $\omega$,
can be considered as a generalized set up of~\cite{Lev95-2}, we will obtain
an accelerating scale factor for the early
Universe, without encountering the above-mentioned problems of~\cite{Lev95-2}.
Moreover, we will show that requirements of an inflationary epoch (specially, the nominal as well as
sufficient conditions) are satisfied in a more convenient manner when the noncommutativity parameter is present.

In this paper, by assuming a spatially flat
FLRW universe, a generalized BD theory, and
introducing a particular NC Poisson bracket
between the BD scalar field and the logarithm of
scale factor, we review the solutions (associated to the NC equations of motion) and discuss the effects of
noncommutativity.

\section{Noncommutative Brans-Dicke Cosmological Setting}
\indent
\label{NC-BDT}
We start with the spatially flat FLRW
metric as the background geometry, namely
\begin{equation}\label{metric1}
ds^{2}=-N^2(t)dt^2+e^{2\alpha(t)}\left(dx^2+dy^2+dz^2\right),
\end{equation}
where $a(t)=e^{\alpha(t)}$ is the scale factor and $N(t)$ is a lapse function.

To get a general set of the field equations, let us start with the
Lagrangian associated to the (generalized\footnote{It has
been recently shown~\cite{RFS11,RFM14} that, instead of adding a scalar potential to BD action, such a scalar potential can be
induced from the geometry of an extra dimension.})
BD theory in the Jordan frame as
\begin{eqnarray}\label{lag1}
{\cal L}[g,\phi]&\!\!\!=\!\!\!&\sqrt{-g}\left[\phi R-
\frac{\omega(\phi)}{\phi}g^{\mu\nu}\nabla_\mu\phi\nabla_\nu\phi-V(\phi)\right]\\\nonumber
&\!\!\!+\!\!\!&\sqrt{-g}{\cal L_{\rm matt}},
\end{eqnarray}
where ${\cal L_{\rm matt}}=16\pi\rho(\alpha)$ is the
Lagrangian density of ordinary matter ($\rho$ is the energy density),
which does not explicitly depend on the BD scalar field $\phi$.
The BD parameter $\omega(\phi)$ is a function of the BD scalar field and varies with the space-time point;
$V(\phi)$ is the scalar potential, the greek indices run from zero to $3$ and $R$ is the Ricci scalar
associated to the metric $g_{\mu\nu}$, whose determinant is denoted by $g$.
Therefore, the Hamiltonian of the model is given by
\begin{eqnarray}\label{Ham-1}
{\cal H}\!\!\!&=&\!\!\!-\frac{Ne^{-3\alpha}}{2\chi\phi}
\left[\frac{\omega(\phi)}{6}P_\alpha^2-\phi^2P_\phi^2+\phi P_\alpha P_\phi\right]\\\nonumber
\!\!\!&+&\!\!\!Ne^{3\alpha}\left(V-16\pi\rho\right),
\end{eqnarray}
where $\chi\equiv2\omega+3$, and $P_\alpha$, $P_\phi$ are the
conjugate momenta associated to $\alpha$ and $\phi$, respectively.
In this work, we work with
the comoving gauge, namely, we set $N=1$.
Instead of the commutative phase space where the Poisson algebra is given by
$\{\alpha,\phi\}=0$, $\{P_\alpha,P_\phi\}=0$,
$\{\alpha,P_\alpha\}=1$ and $\{\phi,P_\phi\}=1$, to investigate the effects of a classical
evolution of the noncommutativity on the
cosmological equations of motion, we presume the
following Poisson commutation relations between
the variables as $\{\alpha,\phi\}=\theta$, $\{P_\alpha,P_\phi\}=0$, $\{\alpha,P_\alpha\}=1$ and $\{\phi,P_\phi\}=1$
where the NC parameter $\theta$ is taken as a constant.
Employing these noncommutation relations and the
Hamiltonian~(\ref{Ham-1}) leads us to
deformed equations of motion as
\begin{eqnarray}\nonumber
\dot{\alpha}\!\!\!&=&\!\!\!-\frac{e^{-3\alpha}}{2\chi\phi}
\left[\frac{\omega(\phi)}{3}P_\alpha+\phi P_\phi\right]\\\nonumber
\!\!\!&-&\!\!\!\frac{\theta e^{-3\alpha}}{2\chi\phi}\left[\frac{1}{6}
\frac{d\omega(\phi)}{d\phi}P_\alpha^2-2\phi P_\phi^2+P_\alpha P_\phi\right]\\
&+&\frac{\theta e^{3\alpha}}{\phi}
\left[V(\phi)+\phi\frac{dV(\phi)}{d\phi}-16\pi\rho\right],\label{NC.H.eq1}\\
\dot{P_\alpha}\!\!&=&\!\!e^{3\alpha}\left[-6V(\phi)+16\pi\left(6\rho+\frac{d \rho}{d \alpha}\right)\right]\label{diff.eq2},\\\nonumber
 \dot{\phi}\!\!&=&\!\!-\frac{e^{-3\alpha}}{2\chi}\left(P_\alpha-2\phi P_\phi\right)\\
&-&\theta e^{3\alpha}\left[6V(\phi)
-16\pi\left(6\rho+\frac{d\rho}{d\alpha}\right)\right].\label{NC.H.eq2}\\\nonumber
\dot{P_\phi}\!\!&=&\!\!\frac{e^{-3\alpha}}{2\chi\phi}
\left[\left(P_\alpha-2\phi P_\phi\right)P_\phi+\frac{1}{6}P_\alpha^2\frac{d\omega(\phi)}{d\phi}\right]\\
 \!\!\!&-&\!\!\!\frac{e^{3\alpha}}{\phi}\left[V(\phi)+\phi\frac{dV(\phi)}{d\phi}-16\pi\rho\right],
 \label{diff.eq4}
\end{eqnarray}
where a dot denotes the differentiation with respect to the
cosmic time and
we have assumed $\phi=\phi(t)$.
We should note that the equations of motion associated to
the momenta $P_{\rm \alpha}$ and $P_{\rm \phi}$, under the proposed
NC deformation, are the same as ones in the commutative case and, obviously,
 in the limit $\theta\rightarrow0$, Eqs.~(\ref{NC.H.eq1}) and (\ref{NC.H.eq2})
are reduced to the corresponding standard commutative equations.

In what follows, we investigate the cosmological implications of this
NC BD setting for a very simple case in which the BD parameter is a
constant and the scalar potential and the ordinary matter are absent.

\section{Gravity-Driven acceleration for cosmological models in the noncommutative BD Setting}
\indent
\label{Vacuum-NC-BD}
As mentioned, we would like to assume a very simple case
in which we set $\omega(\phi)=\omega={\rm constant}$, $\rho=0$ and $V(\phi)=0$.
Therefore, from (\ref{diff.eq2}), we get $\dot{P_\alpha}=0$, which gives a constant
of motion. Thus, we get $P_\alpha=c_1$;
also, Eqs.~(\ref{NC.H.eq1}) and (\ref{diff.eq4}) give $P_{\phi}=c_2^{\pm}\phi^{-1}$ where
$c_1$ and $c_2^{\pm}\neq0$ are the integration constants.
These constants are not independent; by substituting them into the
Hamiltonian constraint, we obtain $c^{\pm}_1=(3|c_2|/\omega)\left[-{\rm sgn}(c_2)\pm\xi\right]$
where  $\xi\equiv\sqrt{\chi/3}$, $\omega\neq0$ and ${\rm sgn}(x)=x/|x|$ is the signum function.
Thus, from (\ref{NC.H.eq2}), $\dot{\phi}$ can be given by
\begin{eqnarray}\label{phi-dot}
\dot{\phi}=-\frac{f^{\pm}}{\xi a^3}\hspace{5mm}{\rm where}\hspace{5mm}
f^{\pm}\equiv\frac{|c_2|}{2\omega}\left[-{\rm sgn}(c_2)\xi\pm1\right].
\end{eqnarray}
By employing the obtained expressions
associated to the momenta and the
integration constants, from Eqs.~(\ref{NC.H.eq1}) and (\ref{NC.H.eq2}) we obtain
\begin{eqnarray}\label{H}
H=h^{\pm}\left(\frac{\dot{\phi}}{\phi}\right)\hspace{5mm}{\rm where} \hspace{5mm}
 h^{\pm}\equiv g^{\pm}+\frac{c_2\theta}{\phi},
\end{eqnarray}
where $H=\dot{a}/a$ is the Hubble constant and $g^{\pm}\equiv-\frac{1}{2}\left[1\pm {\rm sgn}(c_2)\xi\right]$.
We should note that the above equations corresponding with
each sign of\footnote{For simplicity of expressing the
quantities, we will sometimes drop the index $\pm$.} $c_2$
yield two branches for the Hubble parameter.
If we assume an attractive
gravity~\cite{Faraoni.book}, i.e., $\phi>0$,
 the values of $\dot{\phi}$ as well as $h$
 corresponding to each branch\footnote{Following~\cite{BV94,Lev95,Lev95-2}, we will call the branches as follows.
  As in the Einstein
  frame, one of the branches always yields an expanding universe, while
  the other gives a contracting universe.
  Therefore, the solutions correspond to the former, and the
  latter are called the X branch and D branch, respectively.
  In our herein model, when $c_2>0$, X branch and D branch solutions
  correspond to the upper sign and lower sign solutions, respectively.
   When $c_2<0$, we note
    the transformations derived after Eq.~(\ref{duality}).}
   must be specified to get an expansion or contraction universes.
  For instance, in a special case, by setting $c_2>0$ and $\theta=0$,
  we obtain $H=-\frac{\dot{\phi}}{2\phi}(1\pm\xi)$.
  In this case, for $\xi<1$, $H>0$ only when $\dot{\phi}<0$,
   and $H<0$ only when $\dot{\phi}>0$ (for both of the branches).
  While, for $\xi>1$, to have a positive Hubble
  expansion we must choose the upper sign for $\dot{\phi}<0$ and the lower sign for $\dot{\phi}>0$.

  Now, let us investigate a general case. We get the acceleration of the scale factor as
$\frac{\ddot{a}}{a}=-\frac{1}{6\phi}\left[\rho^{(\phi)}+3p^{(\phi)}\right]=-\left(\frac{\dot{\phi}}{\phi}\right)^2
\left(2h^2+h+\frac{c_2\theta}{\phi}\right)$, where the energy
density and pressure associated to the BD scalar field are given by~\cite{RFM14}
  \begin{eqnarray}\label{rho-phi}
 \rho^{(\phi)}\!\!&\equiv&\!\!-T^{0(\phi)}_0=3h^2\left(\frac{\dot{\phi}^2}{\phi}\right),\\
 p^{(\phi)}\!\!&\equiv&\!\!T^{i(\phi)}_i=\left(3h^2+2h+
 \frac{2c_2\theta}{\phi}\right)\left(\frac{\dot{\phi}^2}{\phi}\right),\label{p-phi}
  \end{eqnarray}
 where $i=1,2,3$ (with no sum) and we have employed relations (\ref{phi-dot}) and (\ref{H}).
 Therefore, in order to get an accelerating universe, the following constraint must be satisfied
 \begin{eqnarray}\label{h-con}
 2h^2+h+\frac{c_2\theta}{\phi}<0.
 \end{eqnarray}
 As $\rho^{(\phi)}>0$, relation~(\ref{p-phi}) and constraint~(\ref{h-con})
 dictate that the pressure must take negative values.


From (\ref{H}), we get a relation
between the scale factor and the BD scalar field as
\begin{eqnarray}\label{NC-a-phi}
a(t)=a_i[\phi(t)]^{g}e^{-c_2\theta\phi^{-1}},
\end{eqnarray}
where $a_{\rm i}=e^{\alpha_{\rm i}}$ is an integration
constant, which corresponds to $\alpha$ in a
specific time. Eq.~(\ref{NC-a-phi}) indicates that the NC parameter appears in the
power of an exponential warp factor.
We can easily show that this time-dependent warp factor
appears in the differential equation associated
to $\phi$ [see Eq.~(\ref{NC-diff-phi})] and makes it a very
complicated differential equation, such that we have to solve it numerically instead.
Employing Eqs.~(\ref{NC-a-phi}) and~(\ref{phi-dot}), we get a differential equation for
the BD scalar field as
\begin{eqnarray}\label{NC-diff-phi}
\dot{\phi}\phi^{3g}e^{-3c_2\theta\phi^{-1}}=-\frac{f}{a_i^3\xi},
\end{eqnarray}
where, according to~(\ref{phi-dot}), $f$ depends
on $c_2$ and $\omega$.

In the commutative case, dependent on the value of $\omega$,
we obtain two different types of solutions:
(i) when $g=-1/3$ (or $\omega=-4/3$), which corresponds to both the
lower sign (when $c_2>0$) and the upper sign (when $c_2<0$),
the solutions correspond to the de Sitter-like space
as $a(t)=a_{\rm i}\phi_{\rm i}^{-\frac{1}{3}}e^{mt}$ and $\phi(t)=\phi_{\rm i}e^{-3mt}$,
where $\phi_{\rm i}$ is an integration constant,
and $m\equiv\frac{-|c_2|}{8a_{\rm i}^3}[-{\rm sgn}(c_2)\pm3]$,
ii) for $\omega\neq-4/3$, we obtain the generalized version of the well-known O'Hanlon-Tupper
solution~\cite{o'hanlon-tupper-72-KE95-MW95,Faraoni.book} as
$a(t)=\tilde{a}_{\rm i}(t-t_{\rm ini})^{r_{\pm}}$
and $\phi(t)=\tilde{\phi}_{\rm i}(t-t_{\rm ini})^{s_{\pm}}$
with
\begin{eqnarray}\label{OT-solution-2}\nonumber
\tilde{\phi}_{\rm i}\!\!\!&=&\!\!\!\Bigg\{\frac{\mid c_2\mid}{2a_{\rm i}^3\omega}
\left[{\rm sgn}(c_2)\mp\frac{(\omega+1)}{\xi}\right]\Bigg\}^{s_{\pm}},\\\nonumber
\tilde{a}_{\rm i}\!\!\!&=&\!\!\!a_{\rm i}\tilde{\phi}_{\rm i}^g
=a_{\rm i}\Bigg\{\frac{\mid c_2\mid}{2a_{\rm i}^3\omega}
\left[{\rm sgn}(c_2)\mp\frac{(\omega+1)}{\xi}\right]\Bigg\}^{r_{\pm}},
\end{eqnarray}
where $t_{\rm ini}$ is an integration constant and
the exponents $r_{\pm}$ and $s_{\pm}$ are given by
\begin{eqnarray}\label{r-s}
r_{\pm}&=&\frac{1}{3\omega+4}\left[\omega+1\pm {\rm sgn}(c_2)\xi\right],\\\nonumber
s_{\pm}&=&\frac{1\mp 3{\rm sgn}(c_2)\xi}{3\omega+4}.
\end{eqnarray}

Here, we should explain the role of the parameters present in the model.
For a particular case where $c_2>0$ (or $c_2<0$), we get the solutions
corresponding to $(r_{+},s_{+})$ and $(r_{-},s_{-})$ known as
the fast and slow solutions, respectively~\cite{Faraoni.book}.

By assuming $\omega\neq-4/3$ and redefining $\Phi\equiv-{\rm ln}(G\phi)$
(where $G$ is the gravitational constant), it
has been shown~\cite{V91-GV92-TV92-GPR94} that
there are duality transformations as
\begin{eqnarray}\label{duality}
\alpha&\rightarrow&\left(\frac{3\omega+2}{3\omega+4}\right)
\alpha-2\left(\frac{\omega+1}{3\omega+4}\right)\Phi,\\\nonumber
\Phi&\rightarrow&-\left(\frac{6}{3\omega+4}\right)\alpha-
\left(\frac{3\omega+2}{3\omega+4}\right)\Phi,
\end{eqnarray}
 under which the slow and fast solutions are
 interchanged~\cite{L95-L96}, namely, $(r_{\pm},
 s_{\pm})\longleftrightarrow(r_{\mp},s_{\mp})$.
However, in our model for $\theta=0$ herein, from~(\ref{r-s}),
without considering the duality transformations~(\ref{duality}),
we can see that the sign of the
integration constant $c_2$ is responsible for the
mentioned role, interchanging the lower-upper solutions.
More precisely, under interchanging $c_2>0\leftrightarrow c_2<0$,
 the parameters $c_1$, $g$, and $f$ transform
as $(c_1^{\pm}, f^{\pm}, g^{\pm})\longleftrightarrow (-c_1^{\mp},-f^{\mp},g^{\mp})$, and, consequently, we get $(r_{\pm},s_{\pm})\longleftrightarrow(r_{\mp},s_{\mp})$.
 Also for the general NC case, as seen
from~(\ref{H}), the general duality transformations,
not only depend on the $f$, $g$, and the integration constants $c_1$ and
 $c_2$ but also may depend on the noncommutativity parameter.

In our model, due to presence of the three parameters $c_2$, $\omega$ and $\theta$,
we can obtain a good variety of solutions for X and
D branches, for more detail, see~\cite{RM14}.
Here, we are only interested in summarizing some of the results
of Case Ia, namely, the lower sign with $c_2>0$, $-3/2<\omega<-4/3$ and $\theta<0$:
  (i) For very small negative values for the noncommutative
parameter, we have shown that the scale factor
starts from a singular point at $t=0$ and increases
for both of the commutative and NC cases.
However, it always does not have the same behavior for these cases.
For the commutative case, we always get $\ddot{a}>0$, while for the noncommutative
case, for the very early times, we have $\ddot{a}>0$, but
at the special point, it turns to be negative;
namely, after a very small time, the phase changes
and we obtain a decelerating universe.
(ii) The scalar field always
drops for both of the
commutative and NC cases.
(iii) The larger the integration constant $c_2$, the shorter the
time of the accelerating phase.
(iv) The smaller the $\mid\!\theta\!\mid$, the larger
the slope of $a(t)$, namely $\dot{a}$.
(v) In the late times, by assuming the same initial values for
 the parameters (except $\theta$), $\phi(t)$ tends to zero for both
the commutative and NC cases.
However, for the large values of
the cosmic time, in the commutative case, $\ddot{a}$ never
takes a constant value. While, in the
NC case, we get a zero acceleration epoch.

 \section{Kinetic inflation}
  \label{Kinetic inflation}
 In the previous section, by proposing
 a NC setting for the BD theory, we have mentioned that
how we can overcome the graceful exit problem.
 However, these features alone do not guarantee
 a scenario for the resolution of
 the problems with the standard cosmology.

In this section, we first investigate the nominal condition~\cite{Lev95}
for the acceleration associated to the inflation, namely,
 \begin{eqnarray}\label{NC-inflation-1}
d^{^{\rm Hor}}>H^{-1}, \hspace{5mm} {\rm where} \hspace{5mm} d^{^{\rm Hor}}(t)=a(t)\int_{t_i}^{t}dt'/a'.
\end{eqnarray}
 Then, in the rest of this section, we will study the condition for sufficient inflation.

Employing Eqs.~(\ref{phi-dot}), (\ref{H}), and after some manipulations, we obtain
\begin{eqnarray}\label{NC-inflation-3}
 \frac{d{\rm ln}(a^2\phi)}{dt}=\pm(2h+1)\left(\frac{\dot{\phi}}{\phi}\right).
\end{eqnarray}
Using (\ref{phi-dot}) and integrating over $dt$, we get
\begin{eqnarray}\label{NC-inflation-4}
 d^{^{\rm Hor}}=\frac{a^3\phi}{|f|}-\frac{2c_2}{2g+1} d^{^{\rm NC}}
\end{eqnarray}
up to a constant of integration. In Eq.~(\ref{NC-inflation-4}), we introduced
a NC contribution of distance, $d^{^{\rm NC}}$, as
$d^{^{\rm NC}}\equiv\frac{\theta a}{c_2}\int P'_{\phi'}dt'/a'.$
We expect that $d^{^{\rm NC}}$ can add a positive
value to the $d^{^{\rm Hor}}$ to properly assist in satisfying the
requirement associated to the horizon problem.
In order to compare, by the aid of (\ref{phi-dot}), we rewrite Eq.~(\ref{H}) as
 $H=|f|h/(\xi a^3\phi)$.
Employing this result and (\ref{NC-inflation-4}) in the
nominal condition (\ref{NC-inflation-1})
yields
\begin{eqnarray}\label{Nominal-con}
D^{^{\rm NC}}\equiv d^{^{\rm Hor}}\!\!-H^{-1}=
\frac{\phi a^3}{|f|}\left(\!\!1+\frac{\xi}{h}\right)-\frac{2d^{^{\rm NC}}}{2g+1}>0.
\end{eqnarray}
Obviously, when $\theta$ goes to zero then $d^{^{\rm NC}}$  vanishes, and, thus, the relation
associated to the horizon distance of the commutative case is recovered.
Therefore, the resulted relation is the same as corresponding
one in Ref.~\cite{Lev95} provided that the BD coupling parameter takes constant values within.
Consequently, when $\theta=0$ the only acceptable result
is $0<\xi<1$ ($-3/2<\omega<0$), which is obtained by
choosing either the upper sign for $c_2>0$ or the lower sign for $c_2<0$.

\begin{figure}
\centering\includegraphics[width=3.2in]{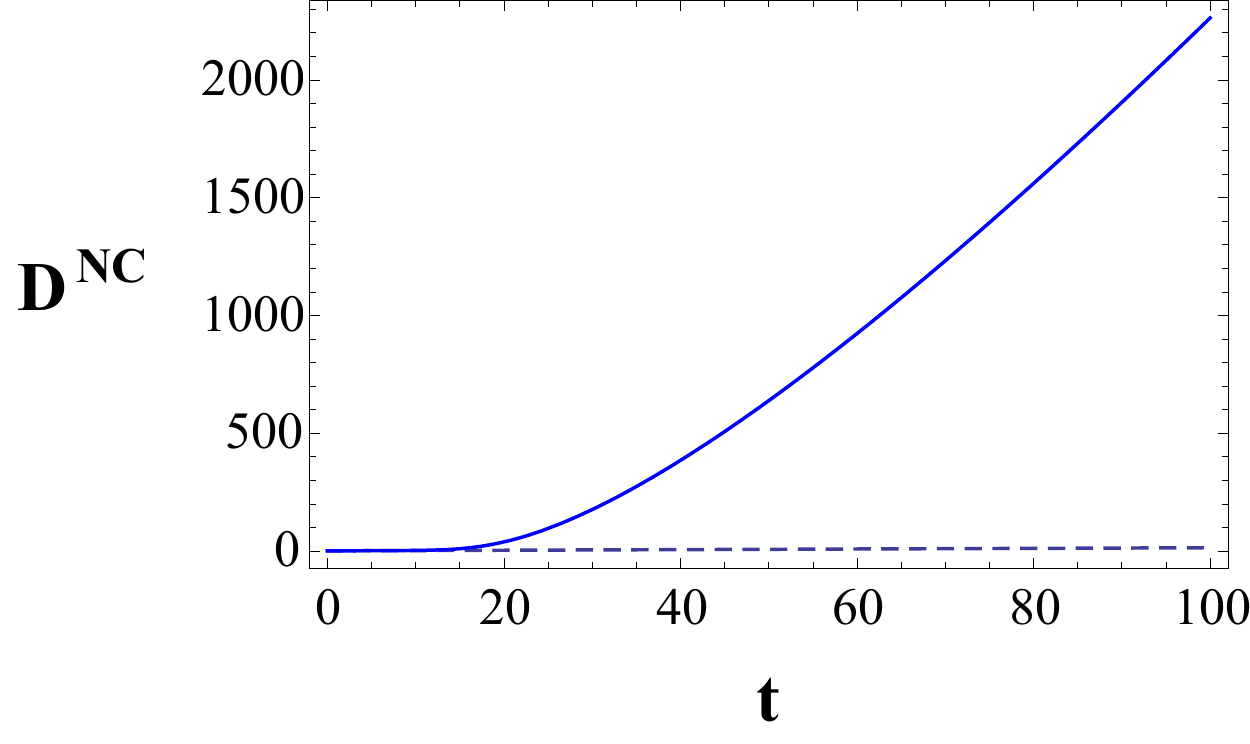}
 \caption{{\footnotesize{{The behavior of the
 $D^{^{\rm NC}}$, the quantity which defined as in (\ref{Nominal-con}), versus cosmic time. The dashed and solid
curves are associated to the commutative and NC cases, respectively.
 This figure is plotted (as an example) to show that the nominal condition~(\ref{NC-inflation-1})
 for the NC solutions (particularly for the case Ia) can be easily satisfied.
The initial values are $\omega=-1.4$, $a_0=1=c_2$, $\theta=0$ (dashed curve) and $\theta=-0.000001$
(solid curve)~\cite{RM14}.}}}}
\label{NR}
\end{figure}

For our herein NC case, it is important to emphasize that
the NC parameter plays a significant role in determining
whether the constraint~(\ref{Nominal-con}) is satisfied or not.
As an example, for both the commutative and the NC (case Ia) cases,
in Fig.~\ref{NR}, we have plotted $D^{^{\rm NC}}$
versus cosmic time. This figure shows that
the constraint~(\ref{Nominal-con}), at all times, for
the NC case can be easier satisfied than
 its corresponding commutative case.

In what follows, we investigate the
condition for sufficient inflation,
which is constrained as~\cite{Lev95-2}
\begin{eqnarray}\label{suf-hor-3}
 \frac{{d}^{^{\rm Hor}}_\star}{a_\star}>\frac{1}{H_0a_0}.
\end{eqnarray}
The quantity appeared in lhs of the above inequality is comoving size of a causally connected region at a
specific earlier time $t_\star$.
 From relation~(\ref{NC-inflation-3}), for
 the specific time $t_\star$, we obtain
 \begin{eqnarray}\label{suf-hor-con-2}
 d^{^{\rm Hor}}_\star=\frac{a^3\phi(1-\delta)}
 {f {\rm sgn}(c_2)}-\frac{2}{2g+1} d^{^{\rm NC}}\Biggr|_\star.
\end{eqnarray}
We should note that the integration constant, which was removed in relation~(\ref{NC-inflation-4}), has now
been included in $\delta\equiv\frac{a_i^2\phi_i}{a^2\phi}$ where the subscript $i$ stands for
initial values. By assuming that $\phi>0$, we always have $\delta\geq0$.
In order to construct the inequality~(\ref{suf-hor-3}) in our herein NC model, which can be
appropriately compared with the corresponding commutative one, we have employed a few assumptions and a lot of manipulations.
Let us here just express the final result~\cite{RM14}:
\begin{eqnarray}\label{suf-hor-p}
 \frac{a_\star^2\phi_\star}{a_{\rm end}^2\phi_{\rm end}}&\gtrsim&
 \left(\frac{M_0}{\sqrt{\bar{\alpha}_0}T_0}\right)(1-\delta_\star)^{-1}\xi^2\\\nonumber
 &\times&\frac{1}{4\pi\epsilon \mid f\mid (g+\frac{c_2\theta}{\phi})^2
 [1-\frac{2d^{^{\rm NC}}_\star}{(2g+1)\mid f\mid(1-\delta_\star)a_\star^3\phi_\star}]},
 \end{eqnarray}
 where the time $t_{\rm end}$ is devoted to the end of inflation
 in which the entropy is produced. The subscript $0$ stand
 for the present epoch, $M_0$ is the value of the Planck mass today
 and $\bar{\alpha}_0=\gamma(t_0)\eta_0=(8\pi/3)(\pi^2/30)\bar{g}(t_0)\eta_0$, in which $\eta_0$ is
for the ratio today of the energy density in matter to that in radiation.
 The constraint~(\ref{suf-hor-p}) is the generalized
 (noncommutative) version of the one resulted
 in~\cite{Lev95-2} provided the BD coupling is taken as a constant therein.

We should note that, in the commutative case, the
constraint~(\ref{suf-hor-3}), even with assuming a favorite set of initial
conditions, implies that the quantity $a^2\phi$ decreases with the cosmic time,
which also indicates that the Planck mass
must decrease during inflation. Consequently, to get
an admissible result, a branch change must be induced~\cite{Lev95-2}.

\begin{figure}
\centering\includegraphics[width=3.2in]{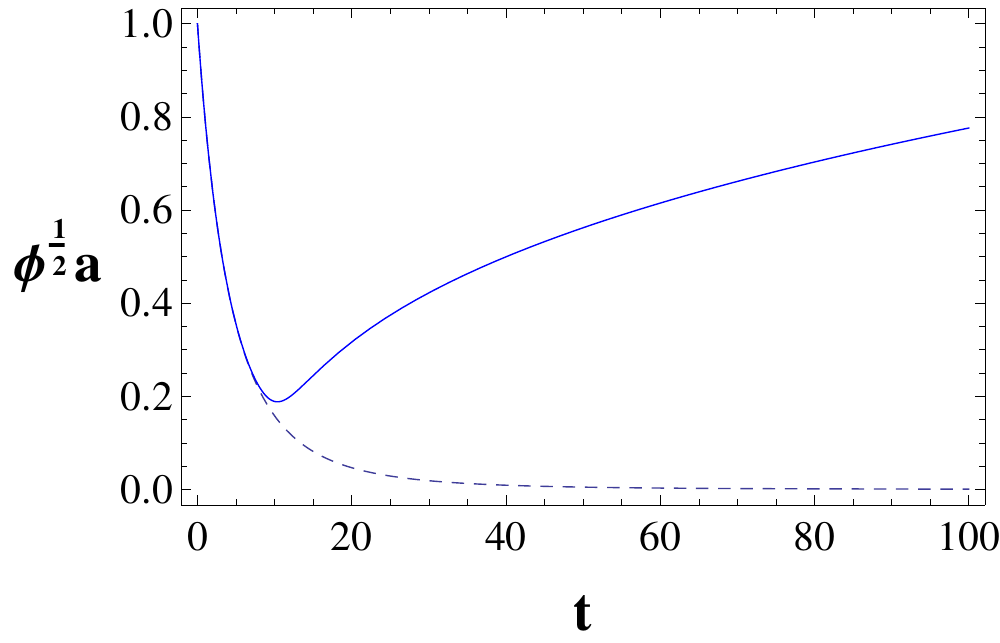}
 \caption{{\footnotesize{{The time behavior of $a\phi^{1/2}$ for the
 commutative case (dashed curve) and NC case (solid curve).
 The initial conditions are $c_2=1=a_0$, $\omega=-1.36$ and $\theta=-0.000001$ for the NC case~\cite{RM14}.}}}}
\label{D-branch}
\end{figure}

In the NC case, considering the
lower sign with $c_2>0$ (D branch) and employing~(\ref{NC-inflation-3}),~(\ref{H})
and the expression for $g^{\pm}$, we obtain $d{\rm ln}(a\phi^{\frac{1}{2}})/dt=
-|f|/(2\xi a^3\phi)\Big[{\rm sgn}(c_2)\xi+\frac{2c_2\theta}{\phi}\Big]$, which, for the commutative case, is reduced to $d{\rm ln}(a\phi^{\frac{1}{2}})/dt=-|f|/(2a^3\phi)$.
  It is easy to show that that, for the commutative case, the
 quantity $a\phi^{1/2}$ always decreases with the cosmic time, see Fig.~\ref{D-branch}.
 We should emphasize that such a result is not in agreement with the observational data~\cite{RM14}.
 For a general NC case, fortunately,
 for $c_2>0$, $\theta<0$, and lower case, i.e., the D branch, we have
 shown numerically that at the early times, $a\phi^{1/2}$ behaves
 similar to its corresponding in the commutative case.
 However, after reaching a nonzero minimum, it starts to increase, see~Fig.~\ref{D-branch}.

\section{conclusions}
\label{Conclusions}
In this paper, we have introduced a NC BD setting.
Such a scenario bears much resemblance to the settings assumed in
NC quantum cosmology
or a few classical noncommutative
gravitational/cosmological investigations in theories
alternative to general relativity~\cite{GSS11,RZMM14,RFK11,RZJM16}).

We have constructed
a NC generalized BD setting to include key ideas of duality and
branch changing as well as gravity-driven acceleration and
kinetic inflation. Then, we have assumed that $N=1$ and there is neither a scalar potential
nor a cosmological constant. Moreover, we have assumed that the Lagrangian
density of the ordinary matter is absent.

In this scenario, we have found that the power-law scale
factor of the Universe is not similar to the commutative case,
but instead it is generalized to be multiplied with a time-dependent exponential warp
factor.
Moreover, we have seen that the BD scalar field is not in the form of
a simple power function of time (similar to its
corresponding commutative case), but instead, it can be obtained from a more
complicated differential equation.

For $\theta=0$,
we have shown that our herein model yields an extended model of the de Sitter--like space
 and O'Hanlon-Tupper solutions.
 In the latter,
the integration constants play the role of the duality transformations
introduced in the context of the BD theory~\cite{L95-L96}.

For case Ia, we have shown that
the scale factor always accelerates in the commutative
case, while, for the NC case, it
accelerates only for very early times, and after a
very short time, it turns to give a decelerated universe.
Consequently, our model can be an appropriate inflationary
model, in which we can overcome the graceful exit problem.
Moreover, contrary to the
commutative model in which the scale factor always accelerates in late times,
we get a zero acceleration epoch for the Universe in the NC case.
It has believed that this result can be interpreted as coarse-grained explanation of the quantum gravity footprint.

We have shown that both the nominal and sufficient requirements associated to the
inflation can be fully satisfied in our NC model.

In the BD theory with a variable $\omega$~\cite{Lev95-2}, for a kinetic inflation model,
all the accelerations in the D branch suffer from the graceful exit problem.
 However, for our NC model, we have shown that his problem is
 appropriately solved due to the presence of the NC parameter: at the very
 early times, although, the same as the commutative case, $a\phi^{1/2}$
 decreases with the cosmic time, while after a very short time, it starts to increase with time, which is
 in agreement with observations.

\section{ACKNOWLEDGMENTS}
S. M. M. Rasouli is
grateful for the support of
Grant No. SFRH/BPD/82479/2011 from the Portuguese
Agency Funda\c c\~{a}o para a Ci\^encia e Tecnologia.
He is also extremely grateful to the organizers of the 16-th Gamow International
Conference/Summer School, specially, prof. Alexander Zhuk, for their warm hospitality during his
attendance at the conference as well as his visiting time.
This research work was supported by Grants No.
CERN/FP/123618/2011 and No. PEst-OE/MAT/UI0212/2014.\\

\end{document}